# A model-independent technique to determine one-dimensional radio source structure from interplanetary scintillation (IPS) observations

*by*

V R Venugopal[1], P K Manoharan[2], D G Banhatti[1], S Edwin Jayaraj[1]

[1]School of Physics, Madurai Kamaraj University, Madurai 625021, India
[2]Radio Astronomy Centre (TIFR), P B 8, Udhagamandalam 643001, India

[The method described was used by S Edwin Jayaraj for his M Phil project under the guidance of the first and third authors at Madurai and the second author at Udhagamandalam.]

(1992)

**Abstract.** We outline a method of deriving one-dimensional phaseless visibility along solar wind direction from interplanetary scintillation power spectrum, together with the known visibility of a calibration source. The method is illustrated briefly. Details may be found in Edwin Jayaraj (1990).

**Introduction**
Observations of the intensity variations of the radio emission from a distant source as the waves pass through the solar wind plasma have provided an inexpensive way to determine the scintillating strength and the one-dimensional angular size of the radio source along the solar wind velocity for about three decades now. The techniques used to derive the structure have always assumed a model for the plasma turbulence in the interplanetary medium. We have developed a new technique (Banhatti et al 1983, Banhatti 1985) for deriving the phaseless visibility of a radio source using the corresponding known function for another source. The phaseless visibility has more information than just the scintillating size and the scintillating flux density, essentially the only two quantities derived hitherto.

**Method**
The IPS power spectrum $P_{extd}(f)$ of an extended source is related to its visibility $B_{extd}(s)$, i.e., the Fourier transform of its brightness distribution, through the power spectrum $P_{pt}(f)$ of a unit point source by (Banhatti et al 1983, Banhatti 1985) (see figure)

$$|B(Z.f/v)|^2 = P_{extd}(f)/P_{pt}(f), \quad \text{or} \quad |B(s)| = \{P_{extd}(v.s/Z)/P_{pt}(v.s/Z)\}^{1/2},$$

Where $Z \approx \cos\varepsilon$ AU is the distance to the scattering screen for an IPS observation at solar elongation $\varepsilon$, while v is the solar wind speed (km/sec) at the screen, and f is the IPS frequency (Hz) from the observed power spectrum plot (or function table). Using this relation for two extended sources having power spectra $P(f)$ & $P_{cal}(f)$ and visibilities $B(s)$ & $B_{cal}(s)$, the point-source power spectrum $P_{pt}(f)$, which depends on the model assumed for the interplanetary medium turbulence, can be eliminated to give

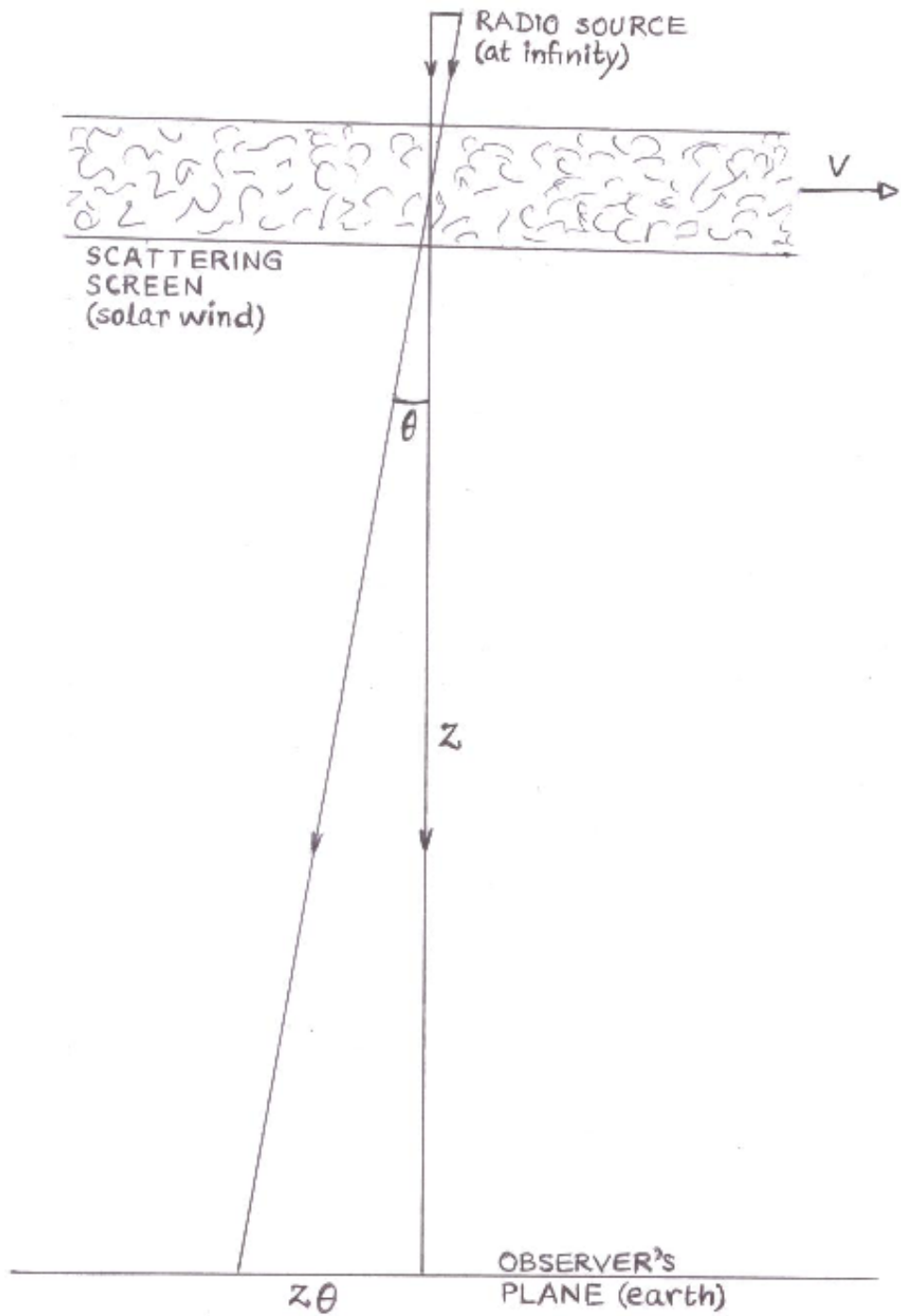

Thin screen model for the scattering of radio waves by the interplanetary medium (i.e., solar wind)

$$| B(s) | = \{ P(f) / P_{cal}(f) \}^{1/2} . | B_{cal}(s) |, \qquad \text{where } s = Z.f / v,$$

relating the desired phaseless visibility of the target source to the known one for the calibration source via the observed power spectra for the target and calibration sources.

**Preliminary Illustration**

As a preliminary illustration of the method, we observed in April and May 1988 the strongly scintillating radio sources 3C48 and 3C119 along with the calibrator CTA21 (size ≈ 50 milliarcsec, scintillating flux density ≈ 5 Jansky). Assuming Gaussian brightness distributions for the sources and a power-law model for the power spectrum of turbulence in the medium, it is possible to fit (Manoharan 1991) for several parameters of the medium and the sizes of the sources (see Table). The method outlined above needs only the solar wind speed v, and gives (Edwin Jayaraj 1990)

$$\theta^2 = \theta_{cal}^2 . (v / v_{cal})^2 + (4.\ln 2 / \pi^2).(K.v^2 / Z^2)$$

for the size $\theta$ of the target source in terms of the slope K of the ratio of the two power spectra plotted against $f^2$. The values of $\theta$ so determined are 240 +/- 25 milliarcsec for 3C48 and 100 +/- 10 milliarcsec for 3C119, comparing favourably with other determinations.

**Table**

| Source | Obsvn date | Power-law index | Elongation ε (deg) | Speed (km/sec) | θ (size) milliarcsec |
|---|---|---|---|---|---|
| 3C48 | 02.04.1988 | -3.0 | 30.3 | 500 | 310 |
| 3C119 | 10.05.1988 | -3.0 | 29.5 | 500 | 100 |
| CTA21(Cal) | 09.04.1988 | -3.3 | 31.0 | 400 | 50 |

-x0x-